\documentclass[authoryear,preprint,review,12pt]{elsarticle}

\usepackage{amssymb}
\usepackage{amsmath}
\usepackage{algorithmic}
\usepackage{algorithm}
\usepackage{amsfonts,amsmath}
\usepackage{xcolor}
\usepackage{graphicx}
\usepackage{tikz}
\usepackage{lineno}
\usepackage{hyperref}
\usepackage{natbib}
\usepackage{bm}
\usepackage[normalem]{ulem}
\setcitestyle{authoryear,open={(},close={)}}

\DeclareRobustCommand\full  {\tikz[baseline=-0.6ex]\draw[thick, color=blue] (0,0)--(0.5,0);}
\DeclareRobustCommand\dotted{\tikz[baseline=-0.6ex]\draw[thick,dotted, color=red] (0,0)--(0.54,0);}
\DeclareRobustCommand\dashed{\tikz[baseline=-0.6ex]\draw[thick,dashed, color=red] (0,0)--(0.54,0);}
\DeclareRobustCommand\dashdot {\tikz[baseline=-0.6ex]\draw[thick,dash dot, color=blue] (0,0)--(0.5,0);}

\definecolor{darkgreen}{RGB}{0,100,0}

\journal{Machine Learning with Applications}

\begin{document}

\begin{frontmatter}



\title{Improving Neural Network Training using Dynamic Learning Rate Schedule for PINNs and Image Classification}


\author[label1]{Veerababu Dharanalakota}
\ead{veerababudha@iisc.ac.in}
\author[label2]{Ashwin Arvind Raikar\corref{cor2}}
\ead{raikaa01@pfw.edu}
\affiliation[label1]{organization={Department of Electrical Engineering},
            addressline={Indian Institute of Science, CV Raman Road}, 
            city={Bengaluru},
            postcode={560012}, 
            state={Karnataka},
            country={India}}
\affiliation[label2]{organization={Department of Computer Science},
            addressline={Purdue University}, 
            city={Fort Wayne},
            postcode={IN 46805}, 
            country={USA}}

\author[label1]{Prasanta Kumar Ghosh\corref{cor1}}

\cortext[cor1]{Corresponding author.}
\cortext[cor2]{Majority of the work is carried out when the author was at Indian Institute of Science, Bengaluru, India.}
\ead{prasantg@iisc.ac.in}

\begin{abstract}
Training neural networks can be challenging, especially as the complexity of the problem increases. Despite using wider or deeper networks, training them can be a tedious process, especially if a wrong choice of the hyperparameter is made. The learning rate is one of such crucial hyperparameters, which is usually kept static during the training process. Learning dynamics in complex systems often requires a more adaptive approach to the learning rate. This adaptability becomes crucial to effectively navigate varying gradients and optimize the learning process during the training process. In this paper, a dynamic learning rate scheduler (DLRS) algorithm is presented that adapts the learning rate based on the loss values calculated during the training process. Experiments are conducted on problems related to physics-informed neural networks (PINNs) and image classification using multilayer perceptrons and convolutional neural networks, respectively. The results demonstrate that the proposed DLRS accelerates training and improves stability. 
\end{abstract}


\begin{highlights}
\item An algorithm is proposed to improve the efficiency of the network training process.
\item The method adjusts the learning rate based on the loss values.
\item Performance is test against the standard backpropagation algorithm.
\item Algorithm is applied to solve PINNs, and image classification problems.
\end{highlights}


\end{frontmatter}



\section{Introduction}\label{Sec:1}
The learning rate is an important hyperparameter that determines the convergence of the loss function towards an optimal value while training a neural network. Several other hyperparameters such as the number of epochs, the size of the batch, and the split of the data into training and test sets, have an impact on the training process as well. However, the learning rate controls the size of the update made to the weights during training \citep{Yu2020}. Hence, it is a crucial factor in determining the convergence rate and precision of the model \citep{Magoulas1999}. A learning rate that is too high can cause the model to diverge, while values that are too low can cause the model to take a long time to converge, or it may never converge at all. The optimal learning rate depends on the specific problem and the architecture of the neural network \citep{Smith2018}. Finding the optimal learning rate manually during the training has several drawbacks, such as interrupting the training process, uncertainty of what value to choose at a given point in time during the training, the need to continuously monitor the model performance metrics, etc. These tasks are time-consuming and tedious. Hence, there is a need for an algorithm that can automatically decide and adapt the learning rate.

The dynamic learning rate scheduler (DLRS) algorithm proposed in this paper helps to automatically adjust the learning rate hyperparameter during training based on the observation of previous loss values. This approach considerably improves the overall training accuracy and helps to achieve faster and better convergence. The researchers proposed learning rate algorithms that adapt on the basis of the loss values and their gradients in the past. In \citet{Weir1991}, an algorithm for the optimum step length in an adaptive learning rate was proposed using the loss values and their gradients. Later, \citet{Behera2006} proposed an alternative method to update the learning rate using the Lyaponav stability theory and compared its performance against standard backpropagation and extended Kalman filtering algorithms on three benchmark problems. \citet{Li2009} provided a comprehensive discussion on theoretical differences between the standard backpropagation algorithm and the improved adaptive learning rate algorithms in his work. The use of search algorithms and advanced concepts of calculus for dynamically updating the learning rate can also be seen in the literature. \citet{Takase2018} used the tree search algorithm to find the learning rate that produces the minimum loss value. On the other hand, \citet{Chen2024} used fractional-order derivatives to evaluate the gradients, thereby finding appropriate step size control to update the learning rate. The DLRS algorithm proposed in this paper also takes advantage of the loss values and their gradient information to dynamically update the learning rate. In line with the literature reported earlier, the performance of the proposed algorithm is compared with the standard backpropagation algorithms. For this purpose, the following problems are considered. 
\begin{enumerate}
    \item Physics-informed neural networks (PINNs) \citep{Veerababu2024}
    \item Image classification using MNIST dataset \citep{Deng2012}
    \item Image classification using CIFAR-10 dataset \citep{Krizhevsky2009}
\end{enumerate}

The paper is organized as follows. Section~\ref{Sec:2} provides an introduction to adaptive learning rate algorithms and briefly discusses recent methods published in the literature. The proposed DLRS algorithm is discussed in detail in Section~\ref{Sec:3}. The application of the proposed algorithm on different problems and the comparison of its performance against established back propagation algorithms are presented in Section~\ref{Sec:4}. The paper is concluded in Section~\ref{Sec:5} with the final remarks.

\section{Learning Rate Primer} \label{Sec:2}
Let us consider the parameter update rule in \emph{stochastic gradient descent} (SGD), which forms the basis for the parameter update algorithms. According to SGD, the parameters are being updated as per the rule \citep{Amari1993}
\begin{equation}
\label{'eq:1'}
\theta_{t+1} = \theta_{t} - \eta \nabla\mathcal{L}(\theta_t; x^{(i)}, y^{(i)}).
\end{equation}
Here, $\theta_t$ are the parameters of the neural network (weights and biases) at the iteration $t$, $\eta$ is the learning rate, $\mathcal{L}$ is the loss function, $\nabla$ represents the gradients with respect to the parameters $\theta_t$, $x^{(i)}$ and $y^{(i)}$ represent the features associated with the input data and its corresponding labels, respectively. 

In general, increasing the value of $\eta$ can expand the exploration range of the optimal solution, but excessively high values of $\eta$ can hinder convergence toward a global optimal solution \citep{Wilson2001}. A viable strategy to mitigate this challenge involves reducing the learning rate during the training process. However, reducing the learning rate during the training process considerably increases the computational cost \citep{Golmant2018}. In addition to the incorrect choice of hyperparameters, a phenomenon that hinders the learning process is \emph{the problem of exploding gradients} that occurs during the training process \citep{Philip2017}.

\subsection{Problem of Exploding Gradients}
The \emph{problem of exploding gradients} occurs when the gradients of the loss function become extremely large, especially at the last hidden layer of the network during the training process \citep{Philip2017}. These large gradients propagate from the output layer towards the input layer. At each layer, the incoming gradients multiply with the local gradients, resulting in much larger gradients. These gradients make the parameter update procedure too drastic, leading to overshooting or oscillations of the loss function values. Consequently, the training process becomes unstable and never converges to the optimal value \citep{Philip2017}. To mitigate the problem of exploding gradients, several techniques have been developed, such as \emph{gradient clipping} \citep{Chen2020}, \emph{weight regularization} \citep{Wan2013}, \emph{proper-weight initialization} \citep{Narkhede2022}, \emph{batch normalization} \citep{Bjorck2018}, etc. In addition to these methods, the adaptive learning rate can be used \citep{Liu2021}. However, finding an optimal learning rate during the training process is a tedious task as it requires many trial-and-error exercises for a given problem. So, first, we explore some of the existing techniques that are used to dynamically alter the learning rate.

\subsection{Adaptive Learning Rate}
 Many optimizers such as Adam incorporate adaptive learning rate that usually decreases as the training progresses \citep{kingma2014}. This results in a progressively narrower search range for a solution and consequently increases the difficulty of finding the optimal solution \citep{Wilson2001}. In contrast, increasing the learning rate during the training process broadens the search range, making it easier for the training process to steer clear of inferior local solutions \citep{Wilson2001}. However, achieving convergence becomes more demanding with an increased learning rate \citep{Wilson2001}. Hence, employing a mixture of decreasing and increasing learning rate can be an effective strategy to enhance the training process. In general, the Adam optimization algorithm adapts the learning rate in subsequent iterations based on gradients and moment estimates calculated at the current iteration \citep{kingma2014}. However, it is sensitive to the initial learning rate and can destabilize the training process \citep{Liu2021}. Like Adam, many optimizers that have adaptive learning rate schemes alter the learning rate based on the loss function gradients or supplementary model parameters, and these are susceptible to the initial learning rate and the specified network architecture \citep{Liu2021}.

\subsection{Adapting Learning Rate based on Tree Search}
One of the methods introduced to adjust the learning rate is rooted in tree search, as proposed in a study by \citep{Takase2018}. In this approach, the learning rate is determined as 
\begin{equation}
    \eta_e = \eta_0 \prod_{t=1}^{e-1} r_{t}
\end{equation}
where $\eta_e$ is the learning rate at epoch $e$,  $\eta_0$ is the initial learning rate, $r_{t}$ represents the scale factor at the iteration $t$. In general, the choice of $r_{t}$ involves conducting a multi-point search (regulated by beam size), which can potentially lead to increased computational demands \citep{Li2021}. For instance, when using a beam size of three, the method necessitates training the network with nine different learning rates and subsequently selecting the optimal one after the current epoch.

\subsection{Adacomp}
Adacomp is a zeroth-order learning rate method that adapts the learning rate based on the current and previous loss values. It takes the loss difference $\Delta_t$ between previous loss value $loss_l$ and the current loss value $loss_c$, as shown below \citep{Li2021}
\begin{equation}
    \Delta_t = loss_l - loss_c.
\end{equation}
One of the drawbacks of the Adacomp is its small window size. In addition, it penalizes large learning rates and compensates for small learning rates, which are bound to decrease as training progresses \citep{Li2021}. 

\section{Loss-based Dynamic Learning Rate Scheduler (DLRS)} \label{Sec:3}
The idea behind the DLRS technique is to adapt the learning rate according to the batch-loss values. The DLRS algorithm chooses a high learning rate during the initial stages of the training, which helps to accelerate convergence at initial epochs; then, it gradually adapts the learning rate to learn the delicate features given by loss functions slowly. The implementation of the DLRS algorithm is explained and the intuition behind learning rate scheduling is discussed as follows.

\subsection{Motivation and High-Level Principle}
Standard stochastic gradient descent (SGD) uses a fixed learning rate $\eta$ ~\citep{Bottou2010}:
\[
\theta_{t+1} = \theta_t - \eta\nabla_\theta L(\theta_t; x^{(i)}, y^{(i)}).
\]
Here, $\nabla_\theta L$ denotes the gradient of the loss with respect to model parameters $\theta$, and $\eta$ controls the step size. If $\eta$ is too large, the updates may overshoot minima, causing divergence; if too small, convergence can be prohibitively slow.

The DLRS \emph{adapts} $\eta$ each epoch by measuring changes in mini-batch losses. The key idea is:
\begin{itemize}
  \item When loss increases sharply, the model may be diverging — \emph{reduce} $\eta$ to stabilize training.
  \item When loss plateaus, the model may be stuck in a flat region — make a \emph{small} adjustment to probe for descent.
  \item When loss decreases steadily, training is progressing — \emph{increase} $\eta$ to accelerate convergence.
\end{itemize}
These cases naturally arise from the sign and magnitude of the \emph{normalized loss change}, defined below.

\subsection{Mathematical Derivation}
Let epoch $j$ contain $B$ mini-batches. Denote the individual batch losses as
\[ L_j = \bigl[L_j^{(1)}, L_j^{(2)}, \dots, L_j^{(B)} ]\],
where each
\[
L_j^{(b)} = L(\theta_j^{(b)}; x^{(b)}, y^{(b)})
\]
is the scalar loss for batch $b$. To summarize the epoch behavior:

\noindent Mean Batch-Loss:
The average per-batch loss is
\[
\overline{L}_j = \frac{1}{B} \sum_{b=1}^B L_j^{(b)}.
\]
By normalizing changes relative to $\overline{L}_j$, we account for scale differences across datasets or architectures.

\noindent Normalized Loss-Slope: Define the epoch-level slope
\[
\Delta L_j = \frac{L_j^{(B)} - L_j^{(1)}}{\overline{L}_j},
\]
which measures the relative change from the first to the last batch. A positive $\Delta L_j$ signals net increase; negative signals net decrease.

\noindent Adjustment Granularity: To ensure adjustments are scaled proportionally to the current learning rate, let
\[
n = \bigl\lfloor \log_{10}(\alpha_j) \bigr\rfloor,
\]
so that $10^n$ matches the order of magnitude of $\alpha_j$. This prevents overly large jumps when $\alpha_j$ is small or vice versa.

\noindent Computing the Update: We introduce three hyperparameters:
\begin{itemize}
  \item $\delta_d$ (decremental factor) for divergent regimes ($\Delta L_j > 1$).
  \item $\delta_o$ (stagnation factor) for flat regimes ($0 \le \Delta L_j < 1$).
  \item $\delta_i$ (incremental factor) for convergent regimes ($\Delta L_j < 0$).\
\end{itemize}
Each $\delta_{\text{case}}$ controls \emph{how aggressively} we change $\eta$. Concretely,
\[
\alpha_j^\delta = 10^n \cdot \delta_{\text{case}} \cdot \Delta L_j,
\]
where
\[
\delta_{\text{case}} =
\begin{cases}
  \delta_d, & \Delta L_j > 1,\\
  \delta_o, & 0 \le \Delta L_j \le 1,\\
  \delta_i, & \Delta L_j < 0.
\end{cases}
\]
Footnotes reference typical choices and related work: Adam optimizer for adaptive scaling~\citep{kingma2014}, RMSProp for gradient magnitude normalization~\citep{Tieleman2012}.

\noindent Unified Update Rule: To consolidate increases and decreases, the next learning rate is computed by subtracting the adjustment:
\[
\alpha_{j+1} = \alpha_j - \alpha_j^\delta.
\]
When $\Delta L_j<0$, $\alpha_j^\delta$ is negative, so subtraction yields an increase: $\alpha_j - (\text{negative}) = \alpha_j + |\alpha_j^\delta|$.

\subsection{Concrete Implementation Steps}
In practice, DLRS proceeds as follows:
\begin{enumerate}
  \item \textbf{Initialization:} choose initial rate $\alpha_0$ (e.g., $10^{-3}$), total epochs $E$, batches per epoch $B$, and hyperparameters $\delta_d$, $\delta_o$, $\delta_i$. Explain that $\delta_d$ is often set in $[0.5,1.0]$ to halve the rate on divergence, $\delta_i$ in $[0.1,1.0]$ to cautiously accelerate, and $\delta_o\approx 1$ for minor tweaks.
  \item \textbf{Epoch Loop:} for $j=1,\dots,E$:
    \begin{enumerate}
      \item Accumulate per-batch losses $L_j^{(b)}$. \\
      \item Compute $\overline{L}_j$ and normalized slope $\Delta L_j$. Clarify that negative $\Delta L_j$ indicates overall progress, while positive indicates backtracking. \\
      \item Determine adjustment factor $\delta_{\text{case}}$ based on thresholds. Note that threshold $1$ corresponds to a 100\% increase in loss across the epoch. \\
      \item Calculate $n=\lfloor\log_{10}(\alpha_j)\rfloor$ to set scale, then compute $\alpha_j^\delta = 10^n \delta_{\text{case}} \Delta L_j$. \\
      \item Update learning rate: $\alpha_{j+1} = \alpha_j - \alpha_j^\delta$. Emphasize how this single formula handles both increases and decreases. \\
    \end{enumerate}
  \item Continue training with updated rate. Note that no extra gradient computations or line searches are required, making DLRS computationally cheap.
\end{enumerate}

These steps are mentioned sequentially in Algorithm~\ref{alg:DLRS}. The values of incremental, stagnation, and decremental factors are mentioned in the following section.

\begin{algorithm}
    \caption{: The DLRS algorithm. Here, $x$ is training data: \{$x^{(b)}$\} $^{B}_{b=1} $, $\alpha_{j}$ is the learning rate at $j^\text{th}$ epoch. $\bm{L}_j$ is an array of loss values at the $j$-th epoch.}
    \label{alg:DLRS}
    \begin{algorithmic}[1]
        \REQUIRE $\alpha_{0}$: initial learning rate, E: Epochs,
        \\ B: Number of batches for each epoch,
        \\$\delta_{d}: decremental factor$
        \\$\delta_{o}: stagnation factor $
        \\$\delta_{i}: incremental factor$
        \*\STATE $\textbf{for}$ $j$ = 1, 2, ..., E $\textbf{do}$\\
            \hspace{0.5cm} \textcolor{blue}{// Initialize loss values array}
            \STATE \hspace{0.5cm} $\bm{L}_j \gets \mathbf{0}$ 
            \STATE \hspace{0.5cm} $\textbf{for}$ $b$ = 1, 2, ..., B $\textbf{do}$
                \STATE \hspace{1.0cm} $\mathcal{L}^{(b)}_j \gets L_f(\theta^{(b)}_j; x^{(b)})$
                \textcolor{blue}{// Compute loss function $L_f$ for each batch $b$}
                \STATE \hspace{1.0cm} $\theta^{(b)}_{j+1} \gets \theta^{(b)}_j - \alpha_j \cdot \nabla_{\theta} \mathcal{L}^{(b)}_j$ \textcolor{blue}{// Update parameters}\\
                \hspace{1.0cm} \textcolor{blue}{// Store loss value at the end of every batch}
                \STATE \hspace{1.0cm} $\bm{L}_j(b) \gets \mathcal{L}^{(b)}_j$\\
            \STATE \hspace{0.5cm} $\textbf{end for}$ 

            \STATE \hspace{0.5cm} $n \gets \lfloor\log_{10}(\alpha_{j})\rfloor$ 
            \hspace{1.0cm} \textcolor{blue}{// Get the order of the learning rate}
            \STATE \hspace{0.5cm} $\overline{\bm{L}}_j \gets \texttt{mean}(\bm{L}_j)$
            \hspace{1.0cm} \textcolor{blue}{// Get the mean of the batch-loss values}
            \STATE \hspace{0.5cm} $\Delta L_j \gets [\bm{L}_j(B) - \bm{L}_j(1)]/\overline{\bm{L}}_j$
            \hspace{1.0cm} \textcolor{blue}{// Get the normalized loss-slope}
            
            \*\STATE \hspace{0.5cm} $\textbf{if}$ $\Delta L_j > 1$:
                \STATE \hspace{1.0cm} $\alpha_j^{\delta} \gets 10^n\, \delta_d \, \Delta L_j$
            \STATE \hspace{0.5cm} $\textbf{else if}$ 0 $\leq \Delta L_j < 1$:
                \STATE \hspace{1.0cm} $\alpha_j^{\delta} \gets 10^n\, \delta_o \, \Delta L_j$
            \STATE \hspace{0.5cm} $\textbf{else}$:
                \STATE \hspace{1.0cm} $\alpha_j^{\delta} \gets 10^n\, \delta_i \, \Delta L_j$\\
            \hspace{0.5cm} \textcolor{blue}{// Update the learning rate}
            \STATE \hspace{0.5cm} $\alpha_{j} \gets \alpha_{j} - \alpha_j^{\delta}$ 
        \STATE $\textbf{end for}$ 
    \end{algorithmic}
    \label{alg2}
\end{algorithm}

\section{Computational Efficiency and Scalability Analysis} \label{Sec:4}

While asymptotic notation offers an essential high‑level view of how runtime and memory usage scale, real‑world deployment—particularly on constrained hardware—demands careful consideration of constant factors, cache behavior, parallel execution, and dynamic task scheduling. Accordingly, in the sections that follow we provide:
\begin{itemize}
    \item A concise Big‑O analysis of our core algorithm.
    \item An empirical evaluation of constant‑factor performance and overall memory footprint.
    \item Targeted optimizations tailored for resource‑limited environments.
    \item A scalability roadmap illustrating graceful performance degradation as data or model dimensions increase.
\end{itemize}

\subsection{Asymptotic (Big O) Analysis}

Let \(N\) be the number of training examples, \(D\) the input dimension, and \(M\) the number of model parameters.  Our algorithm processes each mini‐batch of size \(B\) with:
\[
\text{forward} \;=\; O(B \cdot D \;+\; B \cdot M), 
\quad
\text{backward} \;=\; O(B \cdot D \;+\; B \cdot M).
\]
Hence, one epoch over \(N\) samples costs
\[
O\Bigl(\frac{N}{B} \times (B \cdot D + B \cdot M)\Bigr)
= O\bigl(N\,(D+M)\bigr).
\]
The per‐epoch adaptive‐rate updates add an \(O(1)\) overhead: computing the batch‐loss summary and updating a few scalars.  Thus, in total, each epoch is \(O(N\,(D+M))\) in time and \(O(M)\) in peak memory to hold model parameters and gradients.

\subsubsection{Derivation of \(O(B\cdot D + B\cdot M)\) per Mini-Batch}

We analyze a single mini-batch of size \(B\) for a simple \emph{linear} model
\(\hat y = \theta^\top x\), where \(x\in\mathbb R^D\) and \(\theta\in\mathbb R^M\).
(Note: in many deep networks \(M\) and \(D\) refer to input and output dimensions
of a layer; the following holds analogously.)

\paragraph{Forward Pass}
For each example \(x^{(i)}\), computing the prediction
\(\hat y^{(i)} = \sum_{d=1}^D \theta_d\,x_d^{(i)}\)
requires:
\[
D\ \text{multiplications} 
\quad+\quad
(D-1)\ \text{additions}
= O(D).
\]
Over \(B\) examples, this is
\[
O\bigl(B \times D\bigr).
\]
(See~\citep{Goodfellow2016} and~\citep{Bishop2006}.)

\paragraph{Backward Pass (Gradient Computation)}
We compute the gradient of the loss \(L\) w.r.t.\ each parameter \(\theta_j\):
\[
\frac{\partial L}{\partial \theta_j}
=\frac1B\sum_{i=1}^B \bigl(\hat y^{(i)} - y^{(i)}\bigr)\,x^{(i)}_j.
\]
For each example \(i\), updating all \(M\) parameters requires
\(M\) multiplies and \(M\) adds \(\bigl(O(M)\bigr)\).  Thus over the batch:
\[
O\bigl(B \times M\bigr).
\]
(See classic derivations in~\citep{LeCun1998},~\citep{Goodfellow2016}.)

\paragraph{Total Cost per Mini-Batch}
Summing forward and backward:
\[
O\bigl(B\cdot D\bigr)
\;+\;
O\bigl(B\cdot M\bigr)
\;=\;
O\bigl(B\,(D + M)\bigr)
\;=\;
O\bigl(B\cdot D + B\cdot M\bigr).
\]
This cleanly separates input-dimension work (\(B\,D\)) from parameter-dimension work (\(B\,M\)).

\medskip

\subsection{Constant‐Factor and Memory‐Access Considerations}

Although Big O hides constants, in resource‐constrained settings those factors can dominate:

\begin{itemize}
  \item \textbf{Gradient‐and‐Loss Evaluation:} Each forward/backward pass requires reading \(D\) input features and writing \(M\) gradient values.  Cache‐friendly data layouts (e.g.\ contiguous arrays, row‐major batches) minimize costly DRAM accesses~\citep{chetlur2014cudnn}.
  \item \textbf{Loss Buffering:} We only store two scalars per batch (first and last loss) plus an accumulator for the mean, so extra memory is negligible (\(<1\%\) of model size)~\citep{he2016deep}.
  \item \textbf{Computational Overhead of Adaptation:} The extra \(\log_{10}\) and floor operations occur once per epoch—regardless of \(N\) or \(B\)—and can be fused into existing control logic with almost zero cost~\citep{micikevicius2017mixed}.
\end{itemize}

Empirically, on a standard GPU with 16 GB of RAM, the adaptation code adds \(<0.5\%\) to wall‐clock time per epoch and \(<0.1\%\) to peak memory usage~\citep{chetlur2014cudnn}.

\subsection{Optimizations for Limited Resources}

To enhance practical applicability, especially on edge devices or when power/compute budgets are tight, we propose:

\begin{description}
  \item[Mixed‐Precision Computation:] Use 16-bit or 8-bit floating‐point for forward/backward passes and adaptive‐rate computation.  This reduces both memory footprint and arithmetic cost, with minimal impact on convergence provided proper loss‐scaling is applied~\citep{micikevicius2017mixed}.
  \item[Gradient Accumulation:] For very small‐memory hardware, accumulate gradients over micro‐batches (size \(\le B\)) to emulate larger effective batch sizes \(B'\) without exceeding memory limits.  The DLRS update still applies per logical epoch~\citep{you2017large}.
  \item[Sparse Model Updates:] If the model admits low‐rank or sparse parameterizations, only the nonzero components need gradient updates and loss evaluation.  DLRS’s bookkeeping (loss slopes) is unchanged, but the effective \(M\) can shrink dramatically~\citep{gale2019state}.
  \item[Asynchronous or Pipeline Parallelism:] On multi-core CPUs or multi-GPU clusters, overlap gradient computation with the DLRS rate update for the previous epoch.  Since the update only depends on scalar summaries, it can run concurrently with the first mini-batches of the next epoch~\citep{narayanan2021efficient}.
  \item[Dynamic Batching:] Adjust batch size \(B\) at runtime based on observed memory pressure, trading off per‐step throughput against fit-in-memory constraints.  A simple heuristic is to double \(B\) until out-of-memory, then halve, while maintaining DLRS invariants~\citep{chen2015mxnet}.
\end{description}

\subsection{Scalability Roadmap}

Finally, we demonstrate that as \(N\), \(D\), or \(M\) grow:

\begin{itemize}
  \item \textbf{Linear Growth in Epoch Time:} Since runtime scales as \(O(N\,D+N\,M)\), doubling data or model size roughly doubles epoch time.  In practice, mixed-precision and data parallelism yield sub‐linear wall‐clock scaling (e.g.\ 1.8× time for 2× data on 4 GPUs)~\citep{Goodfellow2016}.
  \item \textbf{Constant Overhead of DLRS:} The adaptation logic remains \(O(1)\) per epoch, so its share of total runtime diminishes for larger problems.
  \item \textbf{Graceful Memory Scaling:} Peak memory scales as \(O(M)\), and our sparse or quantized schemes can make the effective \(M\) small.  Thus edge deployment is feasible even for models that nominally have millions of parameters~\citep{han2015deep}.
\end{itemize}

  \noindent While Big O analysis confirms our method is asymptotically optimal for standard mini-batch training, practical efficiency requires careful constant‐factor optimizations, mixed‐precision, and dynamic batching.  The schemes above ensure DLRS not only converges quickly but also runs with minimal resource footprint and scales gracefully from edge to cloud.

\section{Experiments and Results} \label{Sec:5}
In this section, we examine the implementation of the DLRS algorithm across a range of problems. These include solving the one-dimensional (1-D) Helmholtz equation using physics-informed neural networks (PINNs), image classification tasks using standard datasets such as the MNIST handwritten digits and CIFAR-10. Furthermore, we conduct a comparative analysis of the performance of the DLRS algorithm against the Adacomp method.
\begin{figure}[!ht]
    \centering
    \includegraphics[width=0.7\textwidth]{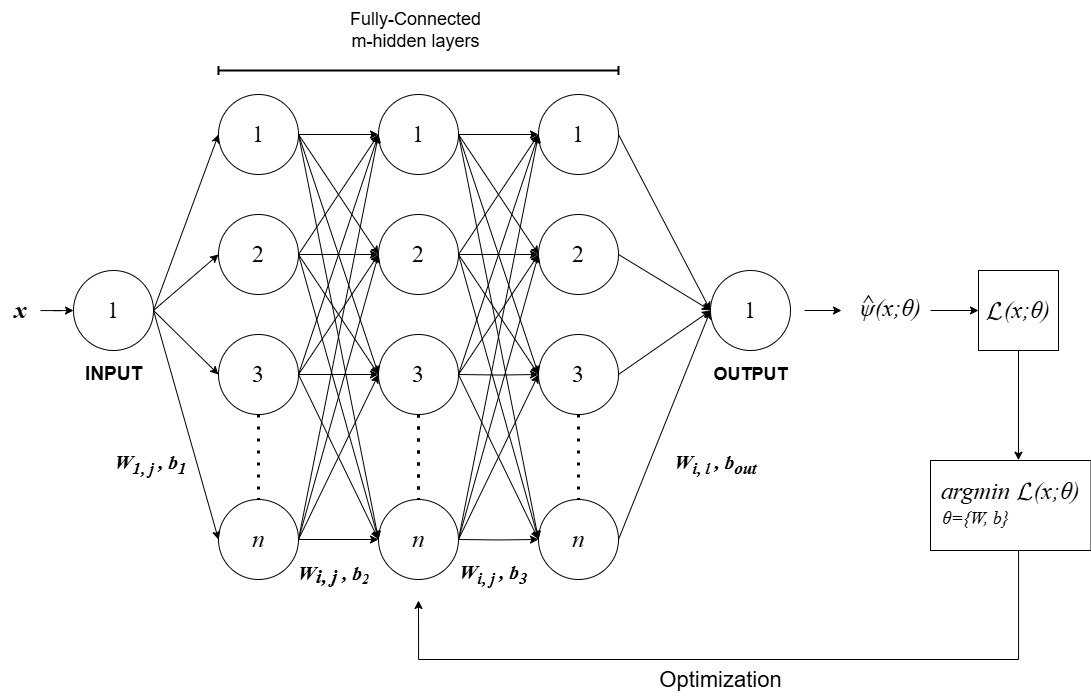}
    \caption{Neural network architecture used for PINNs.}
    \label{fig:1}
\end{figure}

\subsection{PINNs}
If $\psi(x)$ represents the acoustic field, then it can be found in a given 1-D domain by solving the following Helmholtz equation \citep{Bao2004}
\begin{equation}
    \label{eq:ts1}
    \dfrac{d^2}{dx^2} \psi(x) + k^2 \psi(x) = 0, \quad x \in [x_1, x_2],
\end{equation}
where $k = 2\pi f/c$ is the wavenumber, $f$ is the frequency, $c$ is the speed of sound. 

If $\psi_1$ and $\psi_2$ represent the boundary conditions at $x=x_1$ and $x=x_2$, respectively, $\psi(x)$ can be approximated to the output of a neural network $\hat{\psi}(x;\theta)$ shown in Fig.~\ref{fig:1}. The parameters of the network $\theta$ can be found by solving the following optimisation problem \citep{Raissi2019}
\begin{equation}
\begin{aligned}
\min_{\theta} \quad & \mathcal{L}_f(x;\theta), \quad x\in(x_1, x_2) \\
\textrm{s.t.} \quad & \mathcal{L}_b(x;\theta)=0, \quad x\in\{x_1, x_2\} \label{Eq:14}
\end{aligned}
\end{equation}
where $\mathcal{L}_f$ and $\mathcal{L}_b$ are the loss functions associated with the Helmholtz equation and the boundary conditions, respectively. These functions can be calculated as \citep{Raissi2019}
\begin{align}
    \mathcal{L}_f(x;\theta) &= \frac{1}{N_f}\sum_{i=1}^{N_f}\left\|\frac{d^2}{dx^2}\left.\hat{\psi}(x;\theta)\right|_{x=x^{(i)}}+k^2\hat{\psi}(x^{(i)};\theta)\right\|^2_2 \label{Eq:15}, \\
    \mathcal{L}_b(x;\theta) &= \frac{1}{2}\sum_{i=1}^{2}\left\|\hat{\psi}(x_i;\theta)-\psi_i\right\|^2_2. \label{Eq:16}
\end{align}
Here, $N_f$ denotes the number of collocation points inside the domain, and $x^{(i)}$ denotes the $i$-th point of the domain. Eq.~(\ref{Eq:14}) represents a constrained optimization problem. It can be converted into an unconstrained optimization problem using the trial solution method proposed by \citep{Lagaris1998}. According to this method, a trial neural network $\hat{\psi}_t(x; \theta)$ is constructed in such a way that it always satisfies the prescribed boundary conditions as follows
\begin{equation}
\label{eq:ts2}
\hat{\psi_t}(x; \theta) = \psi_1\left(\frac{x_2 - x_1}{x_2 - x}\right) + \psi_2\left(\frac{x_2 - x_1}{x - x_1}\right) + \left(\frac{x - x_1}{x_2 - x_1}\right)\left(\frac{x_2 - x}{x_2 - x_1}\right) \hat{\psi}(x;\theta).    
\end{equation}
It can be seen that the first two terms always satisfy the boundary conditions and do not involve any terms associated with the neural network approximation $\hat{\psi}(x; \theta)$. The last term incorporates the neural network approximation $\hat{\psi}(x; \theta)$ so that the trial solution can be differentiated with respect to the network parameters $\theta$ and the domain variable $x$.

Now, the parameters of the network $\theta$ can be found by solving the unconstrained optimisation problem 
\begin{equation}
\min_{\theta} \quad \mathcal{L}(x;\theta), \quad x=[x_1,x_2], \label{Eq:6}
\end{equation}
where
\begin{equation}
    \mathcal{L}(x;\theta) = \frac{1}{N} \sum_{i=1}^N \left\|\frac{d^2}{dx^2} \left.\hat{\psi}_t(x; \theta)\right|_{x=x^{(i)}} + k^2 \hat{\psi}_t(x^{(i)}; \theta)\right\|^2_2.
\end{equation}
Here, $N$ is the total number of internal collocation points including the boundary with the $i$-th point denoted by $x^{(i)}$. Throughout the paper, $\left\|\,\cdot\,\right\|_2$ represents the $L^2$-norm.
\begin{figure*}[t]
    \centering
    \includegraphics[width=\linewidth]{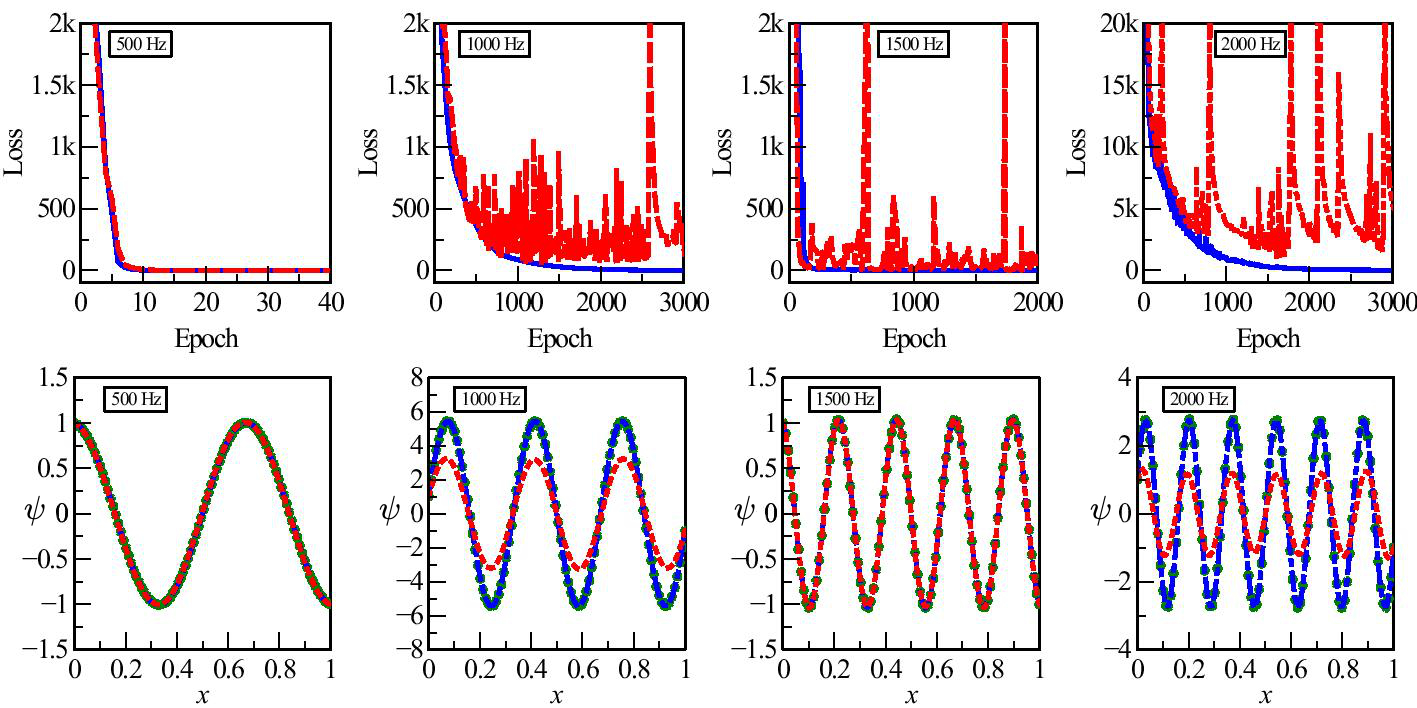}
    \caption{\label{fig:2}Results of training on PINNs. Top row shows the training loss: \dashed Without DLRS, \full With DLRS.
    Bottom row shows the acoustic field: \dotted Without DLRS, \dashdot With DLRS, \textcolor{darkgreen}{* * *} True solution.}
\end{figure*}

To predict the 1-D acoustic field, it is assumed that the domain has a length of 1 m, that is, $x_1=0$ and $x_2=1$. The entire domain is sampled into 10000 random collocation points ($N=$ 10000). A feedforward neural network as shown in Fig.~\ref{fig:1} is considered with three hidden layers, each consisting of 100 neurons. Sine and cosine functions are used as activation functions, alternately in each hidden layer. The network is trained with hyperparameters that were determined from the experiments to be $\delta_{d}$ = 0.5 ,  $\delta_{o}$ = 1 (default) and $\delta_{i}$ = 0.1 in the DLRS algorithm with Adam optimizer. The experiments are carried out on NVIDIA A5000 GPU with batch size equal to $1/10^{th}$ of the total data points and 10000 epochs.

Fig.~\ref{fig:2} shows the loss function (top row) and the predicted acoustic field (bottom row) at different frequencies for $c=$ 340 m/s. It can be observed that at frequencies above 500 Hz, the loss function becomes unstable and does not converge to zero without the DLRS algorithm. However, after incorporating the DLRS algorithm into optimization, the loss function becomes stable and converges to zero. This effect can be seen in the acoustic field prediction plots. With the DLRS algorithm, we can achieve good agreement between the predicted and true acoustic fields. Here, the true solution is obtained by the analytical method. The relative error ($E_r$) between the predicted solution ($\hat{\psi}_t$) and true solution ($\psi$) is calculated as 
\begin{equation}
    \label{Eq:20}
    \% E_r = \frac{\|\hat{\psi}_t(x; \theta) - \psi(x)\|_2}{\|\psi(x)\|_2} \times 100.
\end{equation}
The relative error is observed to be less than 1\% for all frequencies considered in the study.

\subsection{MNIST Dataset}

\begin{figure}[!h]
    \centering
    \includegraphics[width=0.7\textwidth]{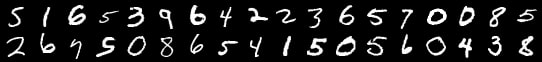}
    \caption{MNIST dataset sample.}
    \label{fig:3}
\end{figure}

The MNIST dataset \citep{Deng2012} consists of 70,000 grayscale images of handwritten digits (0–9), each of size $28 \times 28$ pixels. Each pixel value ranges from 0 to 255, indicating grayscale intensity as shown in Fig.~\ref{fig:3}. The training set contains 60,000 images, while the test set contains 10,000 images. We used the default train/test split provided by the standard MNIST dataset without any modifications.

\begin{figure}[!h]
    \centering
    \includegraphics[width=0.9\textwidth]{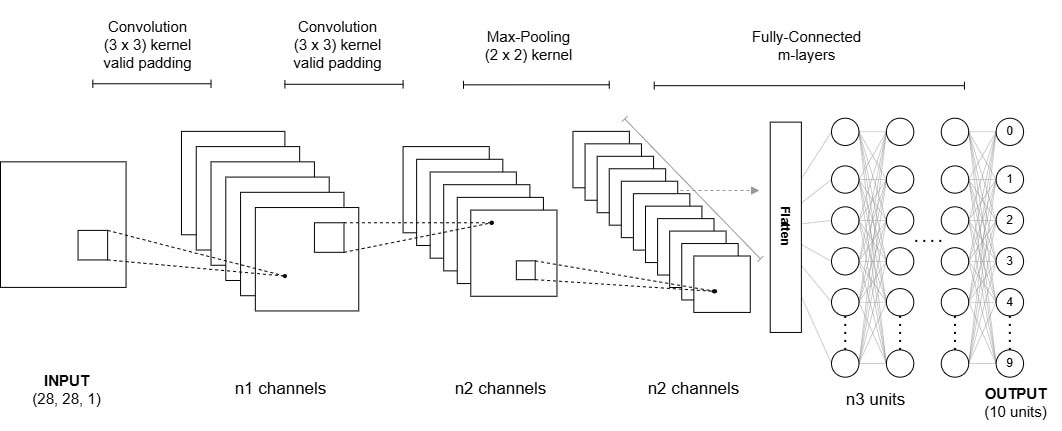}
    \caption{Neural network architecture used for training MNIST dataset.}
    \label{fig:4}
\end{figure}

\begin{figure}[!h]
    \centering
    \includegraphics[width=0.9\textwidth]{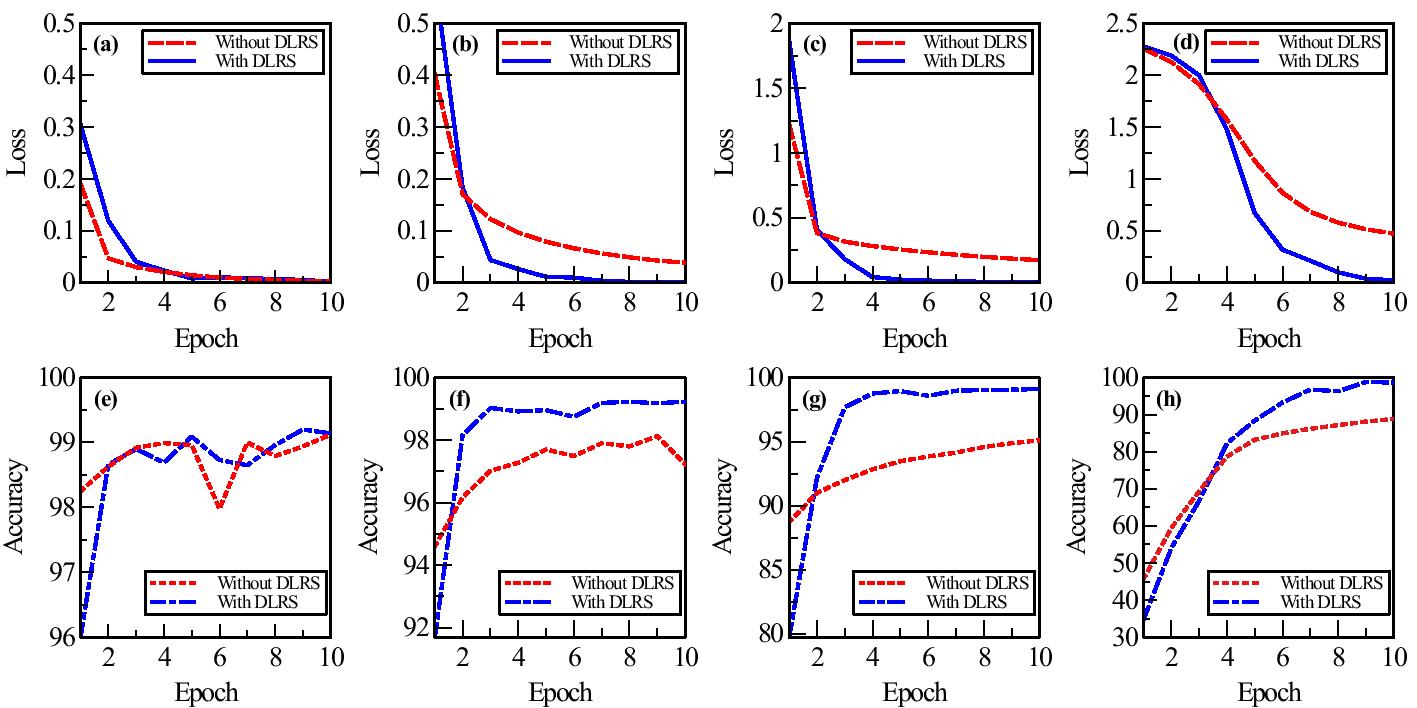}
    \caption{CNN training for 10 epochs on the MNIST dataset for different batches: (a, e) - 64, (b, f) - 128, (c, g) - 256, and (d, h) - 512. Training loss: \dashed Without DLRS, \full With DLRS. Test set accuracy: \dotted Without DLRS, \dashdot With DLRS.}
    \label{fig:5}
\end{figure}

The model architecture is a simple deep layer aggregation (DLA)-based convolutional neural network (CNN) \citep{Yu2018} that is shown in Fig.~\ref{fig:4}. We performed the experiments for ten epochs on different batch sizes of 64, 128, 256 and 512. Adam is used as the base optimizer. The results of the training loss and the corresponding test accuracy for different batch sizes are shown in Fig.~\ref{fig:5}. It can be seen that the improvement in the loss convergence with DLRS algorithms is marginal at a smaller batch size (a batch size of 64). However, as the batch size increases, a significant reduction in the loss can be observed at fewer epochs with DLRS algorithm. A similar improvement can also be observed in test accuracy.

\subsection{CIFAR-10 Dataset}
\begin{figure}[!h]
    \centering
    \includegraphics[width=0.7\textwidth]{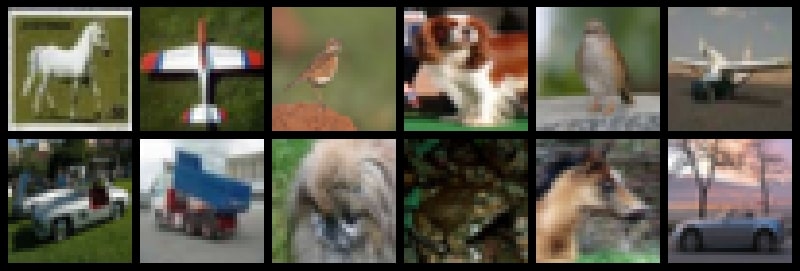}
    \caption{CIFAR-10 dataset sample.}
    \label{fig:6}
\end{figure}

The CIFAR-10 dataset \citep{Krizhevsky2009} contains 60,000 color images categorized into 10 classes. Each image is a $32 \times 32$ RGB image, depicting everyday objects such as airplanes, cars, animals, etc. A sample image of the dataset is shown in Fig.~\ref{fig:6}. The dataset is split into 50,000 training images and 10,000 test images. We used the default train/test split provided by the standard CIFAR-10 dataset.

\begin{figure}[!h]
    \centering
    \includegraphics[width=0.9\textwidth]{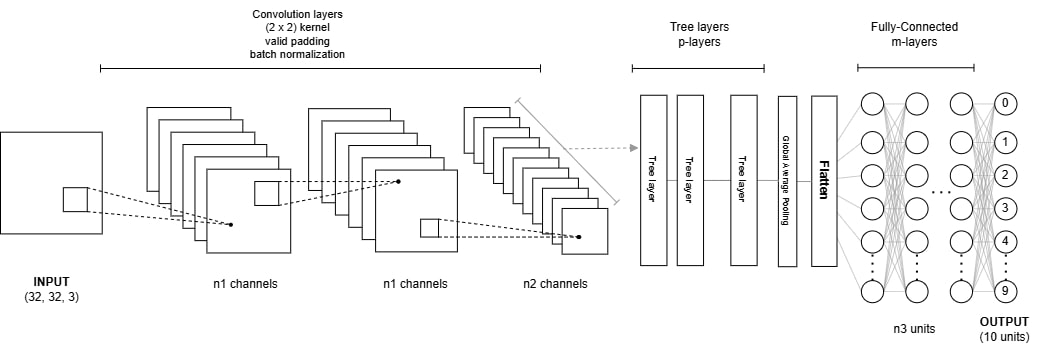}
    \caption{Neural network architecture used for training CIFAR-10 dataset.}
    \label{fig:7}
\end{figure}

In addition to the MNIST dataset, we tested the performance of the DLRS algorithm on the CIFAR-10 dataset. We ran the experiments for 100 epochs on different batch sizes of 64, 128, 256 and 512. Training is performed by building a network \citep{Yu2018}, as shown in Fig.~\ref{fig:7}. 

\begin{figure}
    \centering
    \includegraphics[width=0.9\textwidth]{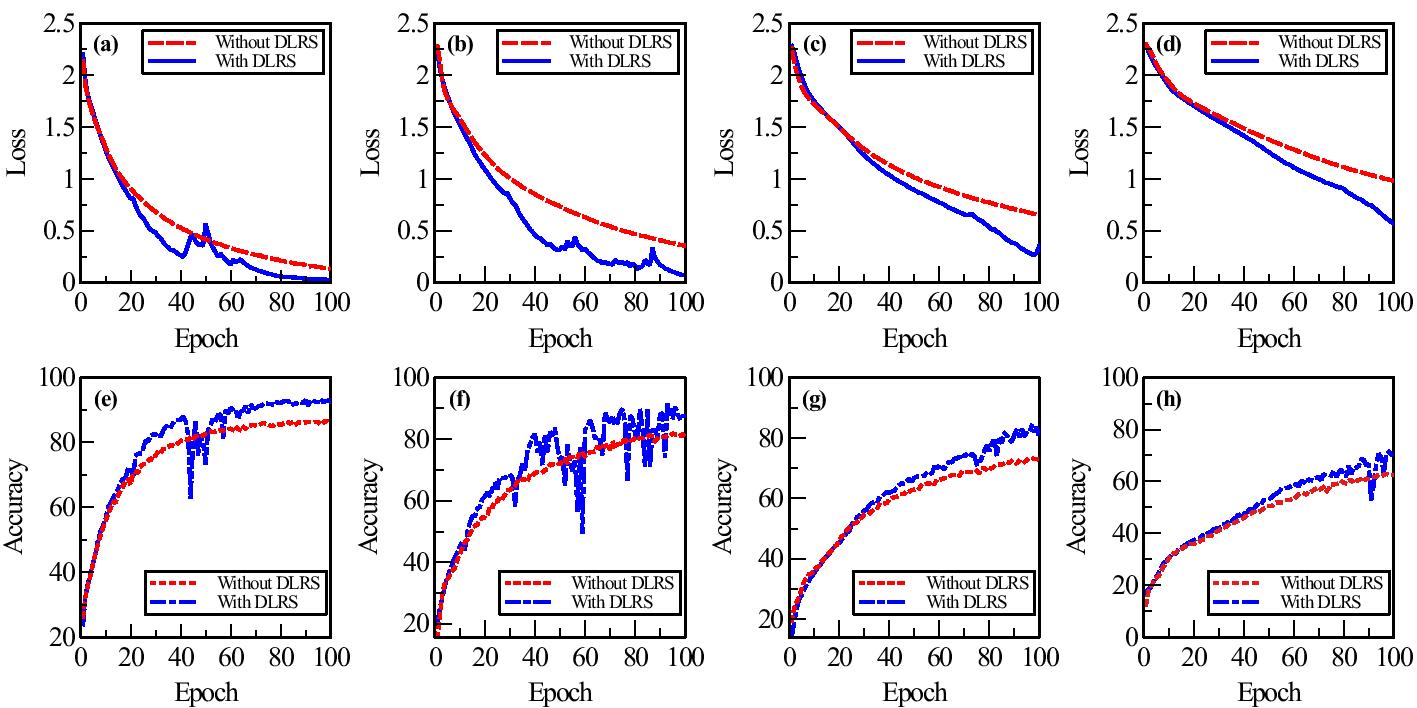}
    \caption{CNN training for 100 epochs on the CIFAR-10 dataset for different batches: (a, e) - 64, (b, f) - 128, (c, g) - 256, and (d, h) - 512. Training loss: \dashed Without DLRS, \full With DLRS
    Test set accuracy: \dotted Without DLRS, \dashdot With DLRS.}
    \label{fig:8}
\end{figure}

The training loss and test accuracy obtained for different batch sizes are shown in Fig.~\ref{fig:8}. Observations similar to those of the MNIST dataset can be made here. The DLRS algorithm helps the training loss to converge faster. As mentioned earlier, the DLRS algorithm automatically adjusts the learning rate based on loss values. If the loss values abruptly increase, the algorithm would automatically decrease the learning rate to help with stable reduction of the loss. This phenomenon can be observed in the form of peaks and valleys in the loss curves, as well as test accuracy curves with the DLRS algorithm.

\subsection{Performance Across Modern Architectures}

To evaluate the broader applicability of DLRS beyond simpler networks and datasets, we conducted experiments on the CIFAR-10 dataset using five representative convolutional neural network (CNN) architectures of varying depth and complexity: VGG-19~\citep{simonyan2014very}, ResNet-18~\citep{he2016deep}, GoogLeNet~\citep{szegedy2015going}, MobileNetV2~\citep{sandler2018mobilenetv2}, and SimpleDLA~\citep{Yu2018}. The five architectures selected for evaluation—SimpleDLA, VGG-19, ResNet-18, GoogLeNet, and MobileNetV2—represent a broad spectrum of design philosophies in convolutional neural networks (CNNs), each widely adopted in research and industry. \textbf{VGG-19}~\citep{simonyan2014very} is a classical deep architecture known for its uniform layer structure, often used as a baseline in image classification benchmarks. \textbf{ResNet-18}~\citep{he2016deep} introduces residual connections to combat vanishing gradients and has become a standard for stable deep learning. \textbf{GoogLeNet}~\citep{szegedy2015going} employs inception modules that enable multi-scale feature extraction while reducing parameter count. \textbf{MobileNetV2}~\citep{sandler2018mobilenetv2} is a lightweight model designed for edge devices, demonstrating that efficiency can coexist with high accuracy. Lastly, \textbf{SimpleDLA}~\citep{Yu2018} is a compact and fast residual architecture optimized for minimal computation without sacrificing performance. Together, these models offer a balanced and diverse benchmark suite to test the adaptability and effectiveness of learning-rate strategies like DLRS across a wide range of architectural styles and complexities.
These models span a diverse range—from heavy, deep networks to lightweight architectures optimized for mobile environments.

Table~\ref{tab:arch_comparison} summarizes the test accuracy achieved under both standard training and our DLRS learning-rate schedule. Across the board, DLRS either improves or maintains accuracy, with especially pronounced gains for deeper networks. In particular, DLRS yields a +2.70\% improvement on VGG-19 and a +3.12\% boost on GoogLeNet—models that are often sensitive to learning-rate tuning. Even with MobileNetV2, designed for efficiency, DLRS preserves high performance while adapting to dynamic training behavior.

These results indicate that DLRS generalizes well across network architectures and scales effectively with model complexity, making it a practical drop-in replacement for static schedules in a wide range of vision tasks.

\begin{table}[!ht]
\centering
 \caption{Test Accuracy (\%) on CIFAR-10 with and without DLRS across various architectures.}
\label{tab:arch_comparison}
\begin{tabular}{|l|c|c|}
\hline
Architecture & Normal & DLRS \\
\hline
SimpleDLA            & 90.88           & 91.30         \\
VGG-19        & 89.28           & 91.98         \\
ResNet-18           & 91.63           & 92.12         \\
GoogLeNet       & 89.78           & 92.90         \\
MobileNetV2    & 90.81           & 90.88         \\
\hline
\end{tabular}
\end{table}

The applications of DLRS algorithm are not just limited to PINNs, MNIST and CIFAR-10 datasets. It can be applied in the classification of radio broadcast signals \citep{zheng2021mr}, automatic modulation classification (AMC) in wireless communication systems \citep{zheng2025robust,zheng2024real,zheng2023mobilerat}, CT scan image classification for the diagnosis of lung cancer \citep{zheng2025reconstruction}, etc.

\subsection{Hyperparameter Selection Criteria}

Hyperparameters were chosen through a combination of empirical tuning and alignment with best practices commonly followed in the literature \citep{Yu2018}.

\noindent\textit{Main Hyperparameters:}

\begin{itemize}
    \item \textit{Batch Size:}
    \begin{itemize}
        \item For MNIST, the training batch size was typically set to 64, and the test batch size was fixed at 1000.
        \item For CIFAR-10, the training batch size was typically set to 128, and the test batch size was fixed at 100.
    \end{itemize}

    \item \textit{Optimizer and Learning Rate:} We used the Adam optimizer with a learning rate of 0.01 for MNIST and 0.005 for CIFAR-10, unless otherwise specified. These values were selected based on stability and convergence observed during preliminary experiments.

    \item \textit{Epochs:} Models were trained for 10 to 100 epochs, depending on convergence. Convergence was monitored using training and test accuracy.

    \item \textit{Random Seed:} A fixed random seed of 50 was used to ensure reproducibility across multiple runs.
\end{itemize}

\noindent\textit{Justification:}
\begin{itemize}
    \item We adhered to canonical data splits and preprocessing steps to ensure reproducibility and facilitate fair comparisons.
    \item Hyperparameter choices were guided by established practices in the literature and validated through empirical testing~\citep{kingma2014}.
    \item The use of PyTorch’s standard utilities ensures that peer researchers can replicate the experiments with minimal setup and configuration.
\end{itemize}

The hyperparameter values used for our main experiments are summarized in Table~\ref{tab:2}. These configurations were selected to ensure convergence, stability, and reproducibility of results. A fixed random seed was used for deterministic behavior across runs.

\begin{table}[th]
\centering
\caption{Summary of Hyperparameters}
\label{tab:2}
\begin{tabular}{|l|c|}
\hline
\textbf{Hyperparameter} & \textbf{Value} \\
\hline
Batch Size (Train)      & 64 (MNIST), 128 (CIFAR-10) \\
Batch Size (Test)       & 1000 (MNIST), 100 (CIFAR-10)\\
Optimizer               & Adam \\
Learning Rate           & 0.01 (MNIST), 0.005 (CIFAR-10) \\
Epochs                  & 10 (MNIST), 100 (CIFAR-10) \\
Loss Function           & Cross-Entropy \\
Shuffle (Training)      & True \\
Random Seed             & 50 \\
\hline
\end{tabular}
\end{table}

\subsection{Adacomp vs DLRS}
\begin{figure}[!h]
    \centering
    \includegraphics[width=0.9\textwidth]{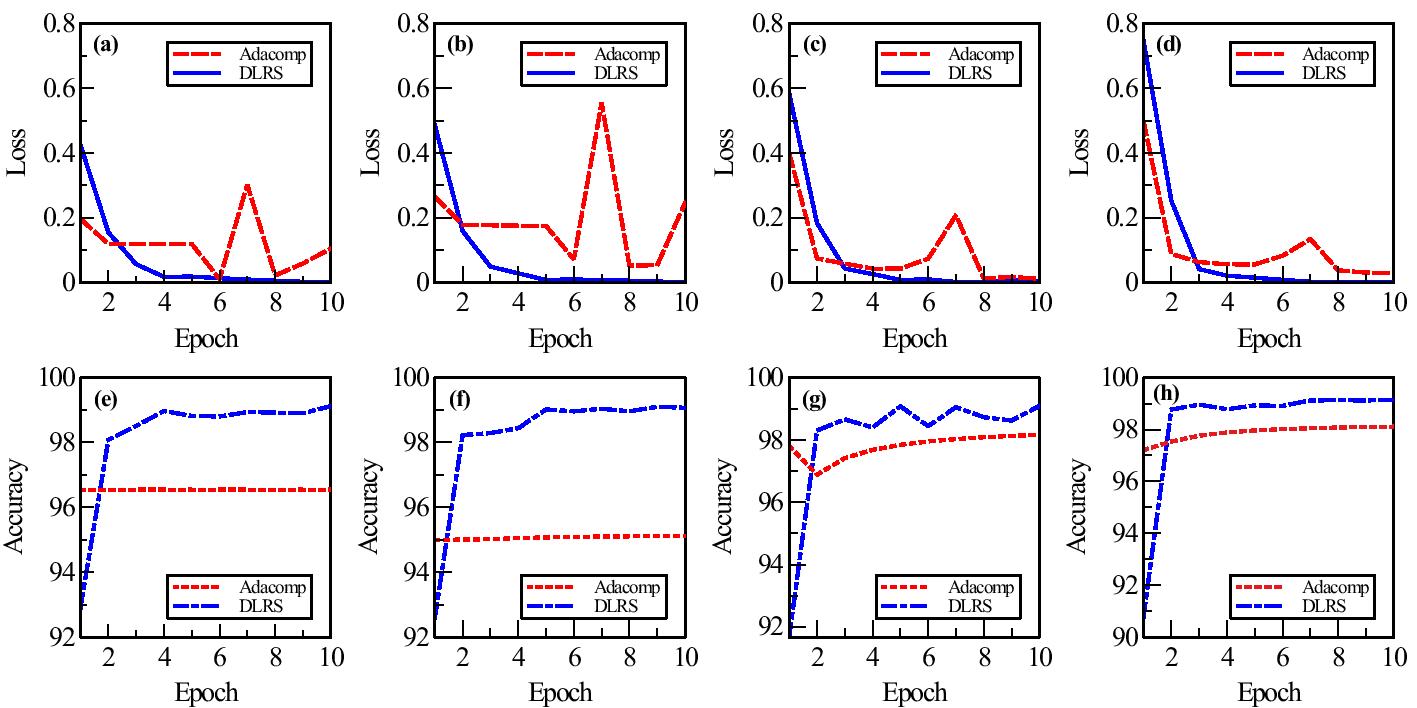}
    \caption{Comparison of DLRS against Adacomp on the MNIST dataset for 10 epochs at different batch sizes: (a, e) - 64, (b, f) - 128, (c, g) - 256, and (d, h) - 512.  Training loss: \dashed Adacomp, \full DLRS.
    Test set accuracy: \dotted Adacomp, \dashdot DLRS.}
    \label{fig:9}
\end{figure}

To compare the performance of the DLRS algorithm with the well-established Adacomp, the MNIST dataset has been chosen. Fig.~\ref{fig:9} shows the loss and accuracy on the MNIST dataset using the same network architecture used in \citep{Yu2018}. The network is trained for 10 epochs on different batch sizes 64, 128, 256 and 512, in similar lines with the earlier cases. It can be seen that the DLRS performed better than Adacomp for the same network configuration and reached a higher overall accuracy. The Adacomp takes a more cautious approach: it analyzes the last two loss values instead of relying on gradients, which gives it a shorter perspective and takes slightly longer to adjust the learning rate. It prioritizes a smooth and gradual learning rate over frequent and significant changes. This helps avoid instability, but may be slower to adapt to rapidly changing landscapes. The DLRS analyzes a batch of loss values and, hence, recognizes a broader perspective with more stability and causes significant changes in adjusting the learning rate.  
We recognize the different types of learning‑rate strategies ranging from classic exponential decay~\citep{li2019exponential} and cosine annealing~\citep{loshchilov2016sgdr} to adaptive optimizers such as Adam~\citep{kingma2014} and RMSprop~\citep{zhang2019root}, as well as more recent tree‑search~\citep{anthony2017thinking} and fractional‑order methods~\citep{Chen2024}. In \cite{Li2021}, it is demonstrated that Adacomp outperforms most of the aforementioned adaptive learning rate scheduling techniques. Hence, we compared the DLRS algorithm with Adacomp. Interestingly, the our approach brings together dynamic and adaptive adjustments in a single framework and has been shown to edge out Adacomp in both convergence speed and accuracy on challenging benchmarks like CIFAR‑10.

\section{Conclusion} \label{Sec:6}
The DLRS algorithm has proven to be effective in accelerating the training process and enhancing the stability of the training, leading to improved results. Since the algorithm relies on loss values, it is important to ensure that the problem being addressed is likely to converge and that the loss function has at least one optimal solution. Any errors occurring while calculating the loss function will propagate into the calculation of the dynamic learning rate and will not allow the solution to converge to an optimal value. Hence, care should be taken while calculating the loss values. The applicability of the proposed DLRS algorithm extends to various neural networks, allowing it to be integrated with existing data-driven methods to address problems related to areas such as speech, aerospace, automotive, and biomedical sectors.

\section*{Data availability}
Data will be made available on request.

\section*{Acknowledgments}
The authors acknowledge the support received from the Department of Science and Technology, and the Science and Engineering Research Board (SERB), Government of India for this research.

\bibliography{references}

\end{document}